\documentclass[aps,prb,twocolumn,amssymb,amsmath,superscriptaddress]{revtex4-1}
\usepackage{bm}
\usepackage{times}
\usepackage{graphicx}
\usepackage{color}
\usepackage{dcolumn}
\usepackage[colorlinks=true, letterpaper=true, pdfstartview=FitV, linkcolor=blue, citecolor=blue, urlcolor=blue]{hyperref}

\begin{document}
\title{High-efficiency photoelectric detector based on monolayer black phosphorus PN homojunction}

\author{Xueying Zuo$^{1}$, Jingjing Cheng$^{1}$, Yulin Liang$^{1}$, Fuming Xu$^{2,*}$, and Yanxia Xing$^{1,\dagger}$ }
\affiliation{
$^1$ Center for Quantum Physics, Key Laboratory of Advanced Optoelectronic Quantum Architecture and Measurement (MOE), School of Physics, Beijing Institute of Technology, Beijing 100081, China \\
$^2$ College of Physics and Optoelectronic Engineering, Shenzhen University, Shenzhen 518060, China }
\begin{abstract}
We numerically investigate the high-efficiency photovoltaic effect in lateral PN homojunction based on monolayer black phosphorus (MBP) by using the non-equilibrium Green's function combined with the density functional theory. Due to the built-in electric field of the PN junction and the wrinkle structure of MBP, the photocurrent excited by either linearly or elliptically polarized light is significantly enhanced in a wide photon energy range. Moreover, because of the electron-photon interaction, the photocurrent is related to atomic orbitals through the polarizing angle of polarized light. Therefore, we can read the orbital information of the band structure from the polarizing angular distribution of photocurrent. These findings suggest the promising application of MBP-based PN homojunction in high-efficiency photoelectric devices and orbital-resolved photovoltaic detection.
\end{abstract}
\maketitle

\section{introduction}
The photovoltaic effect (PVE) converts light energy into electric energy. Since the conversion is clean and sustainable, PVE has been widely applied in the new energy industry. It is found that when irradiated by polarized light (PL), PVE occurs in materials without spatial inversion symmetry (SIS).\cite{Gan2000,Gan2001,Gan2002,Ma2013} PN junctions contain built-in electric fields that naturally break SIS. When photons excite the junction, electron-hole (e-h) pairs are generated. Due to the presence of the built-in electric field, electrons and holes from e-h pairs move in opposite directions and induce a photocurrent.\cite{Ste2004} Hence the PN junction can significantly enhance the photovoltaic efficiency and serves as the foundation of photoelectric devices.

Conventional PN junctions are mostly based on bulk semiconductors.\citep{Riordan1997} With the development of material science, two-dimensional (2D) materials with triangular lattice, such as graphene and transition metal dichalcogenides, offer great potential in 2D PN junctions owing to their unique dispersion properties.\citep{Chen2012,Choi2014,Jin2015,Wang2016,Sun2018,Wu2020,LZhang2020,Gho2021,Zeng2021,Xu2022,Zheng2023} There are typically two configurations for the 2D PN junctions: vertical and lateral. The vertical junction is stacked by weak van der Waals forces, while the lateral configuration is interconnected by strong in-plane covalent bonding between P- and N-type atoms, which is more popular. Lateral PN junctions then can be categorized into homojunction and heterojunction, depending on whether the materials of their P and N regions are the same. Compared with the heterojunction composed of different materials, the homojunction avoids the dislocation introduced by lattice mismatch at the interface. Additionally, because of continuous band alignments, the interface of a homojunction possesses less carrier trap sites.\citep{S2007} Therefore, 2D lateral PN homojunction is a promising PVE platform.

As a 2D material, monolayer black phosphorus (MBP) large surface area-to-volume ratio, which enables efficient light absorption and enhances the generation of electron-hole pairs. MBP holds appealing electrical and optical properties, such as high charge-carrier mobility,\citep{Samuel2014,Li2014,Qiao2014,LZhang14} strong in-plane anisotropy,\citep{Qiao2014,Xia2014,Wang2015,Tran2014} and tunable bandgap.\citep{Qiao2014,Wei2014,Rud2014,Guan2014,Wang2015a,Tran2014} MBP-based photodetectors have broadband photoresponse,\citep{Xie2021,Yuan2015,Guo2021} high polarization sensitivity,\citep{Yuan2015,Guo2021,Li2018} and low power consumption.\citep{Jeon2019,Luo2019,Zhao2020,Xie2021} Hence MBP has been widely used to design various optoelectronic devices, including  photodetector,\citep{Young2017,Jeon2019,Yuan2015,Guo2021,Li2018,Luo2019,Zhao2020,Xie2021} battery,\citep{Xie2018} and field-effect transistor.\citep{Li2014,Xia2014,Wu2015,Buscema2014,Bus2014} Benefiting from recent experimental breakthrough in the doping engineering of MBP,\citep{Xu2016,Lv2018,Yu2020,Jala2020,Hu2020} lateral PN homojunction based on MBP becomes possible and is expected to exhibit efficient photovoltaic response.

In this work, we propose a high-efficiency photoelectric detector based on MBP. Using the nonequilibrium Green's function combined with the density functional theory (NEGF-DFT), we numerically investigate PVE related properties of a 2D lateral MBP PN homojunction. It is found that the built-in electric field of PN junctions greatly enhances the excited photocurrents. Moveover, the wrinkle structure of MBP causes that its $p_{x/y}$ orbital contributes widely to the whole band structure, which significantly improves the PVE efficiency since in-plane photocurrent is determined by the extended states in the $x$-$y$ plane. Due to the electron-photon (e-p) interaction, the induced photocurrent is related to electron momentum $P_{x/y}$ through the polarizing angle of incident light. It is found that momentum $P_{x/y}$ of the photocurrent favors contribution from the extended $p_{x/y}$ orbital of the underlying system. Therefore, we can read the orbital information in the band structure according to polarizing angular distribution of the photocurrent.

The rest of the paper is organized as follows. In Sec.{\color{blue}II}, the model system based on a MBP PN homojunction and numerical methods are introduced. Sec.{\color{blue}III} contains simulation results and relevant discussion on the photoelectric response properties of the model system. Finally, a brief summary is presented in Sec.{\color{blue}IV}.

\section{Model and Method}

\begin{figure}[tbp]
\centering
\includegraphics[width=\columnwidth]{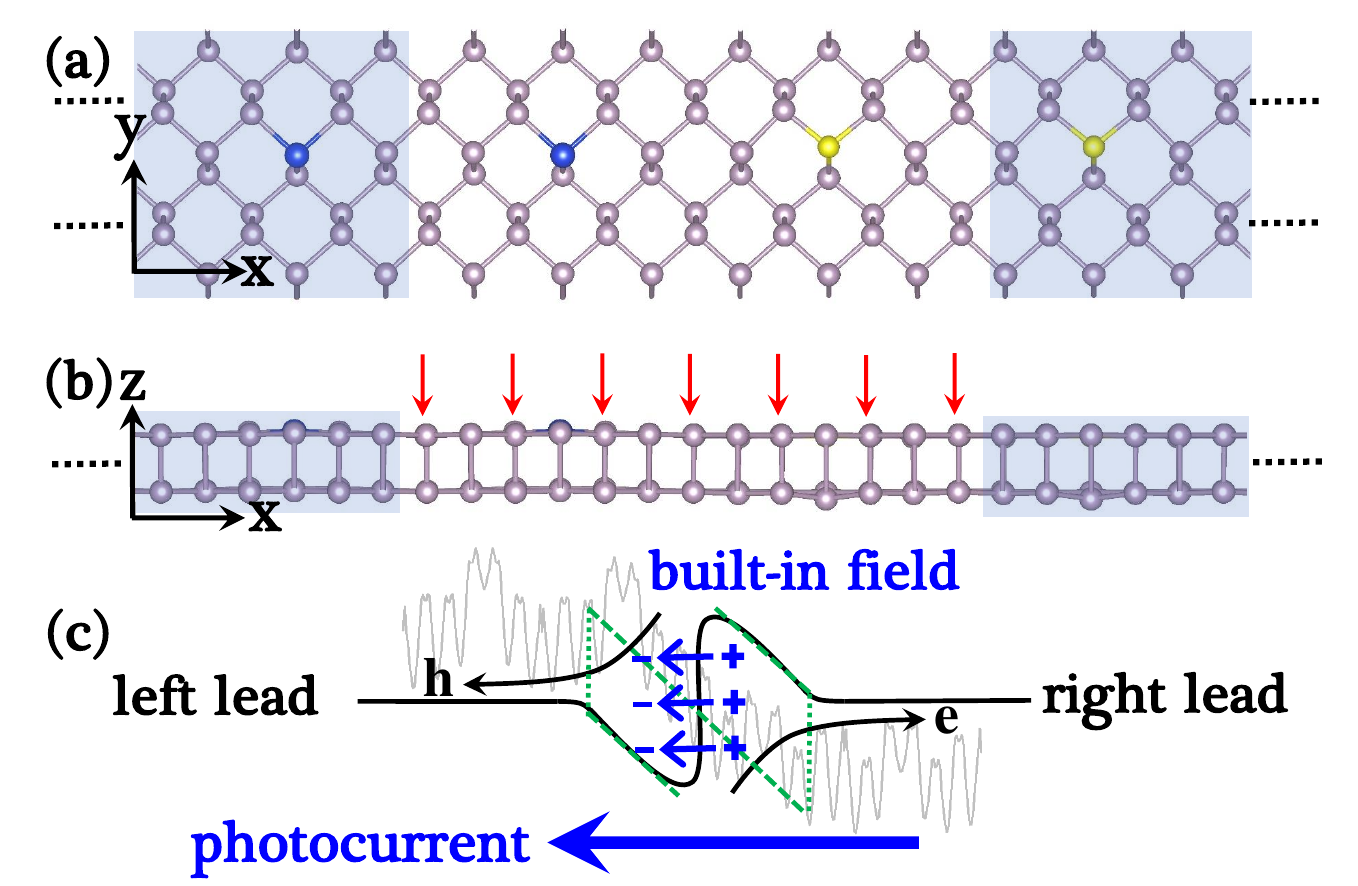}
\caption{(a) Top view of the 2D PN homojunction based on MBP. The blue and yellow spheres denote Si and S atoms, respectively. The shaded parts denotes the supercells which infinitely extend to the $\pm$ $x$-direction. The infinite stripe has periodic boundary conditions in the $y$- and $z$-directions. (b) Side view of the PN homojunction. Polarized lights (red arrows) are projected on the central scattering region.  (c) Schematic diagram of the PVE device. The light gray curve denotes the effective potential energy calculated through nanodcal. The black line labels the effective Fermi level with is curved near the PN junction due to the coupling between the P and N regions. The green lines indicate the potential decline which leads to the charge accumulation (blue + -) and hence the built-in electric field (blue thin arrows). Driven by the built-in field, the photocurrent flows from the right to the left leads (blue block arrow). }\label{fig1}
\end{figure}

First, we illustrate the geometric structure and schematic of the PVE device: a 2D lateral PN homojunction based on MBP. As shown in Fig.\ref{fig1}(a), the PN homojunction is formed by chemically doping Si (blue atoms) and S (yellow atoms) in different regions of MBP, which extends to $x=\pm\infty$. Here atom substitution doping is considered. In the $z$-direction, vacuum regions are maintained larger than 15 \AA\ in order to avoid spurious interactions. Periodic boundary conditions are applied in the $y$- and $z$-directions. Next, geometric optimization is conducted using the Vienna $Ab$ $initio$ Simulation Package (VASP) based on DFT.\citep{Kre1996,Kre1996a} Generalized gradient approximation (GGA) with the Perdew-Burke-Ernzerhof (PBE)\citep{Per1996} parametrization is selected as the exchange-correlation functional. Projector augmented wave (PAW) method is utilized to simulate electron-ion interactions.\citep{Bloechl1994,Kre1999} The plane wave cutoff energy is set as 500 eV. The structure is fully relaxed until the force on each atom is less than 0.01 eV/\AA. Since the atoms of MBP are not in one plane, we also show the side view of the MBP-based device in Fig.\ref{fig1}(b).

The device is divided into three regions: the central region and the semi-infinite left/right leads. The central region is vertically irradiated by PL (the red arrows). The Fermi energy of the P region is lower than that of the N region due to the atom doping. However, when the P and N regions are coupled with each other, the Fermi energy of the P/N region increases/decreases through the effective potential energy of the central region and forms the curved Fermi level near the PN junction. The effective potential energy along the $x$-direction is plotted as the background (the light gray line). As shown in Fig.\ref{fig1}(c), the curved Fermi level leads to the built-in electric field near the joint of the PN junction. From the potential decline (the green dash line), we can estimate the acting range and the strength of the built-in field. While excited by photons, the $e$-$h$ pairs offer effective transport carriers. Driven by the built-in electric field (the blue arrows), electrons/holes flow into the right/left leads and generate a photocurrent.

Subsequently, we calculate the photocurrent using the first-principles quantum transport package, $Nanodcal$, which is based on NEGF-DFT.\citep{Taylor2001,Taylor2001a,hzw} Norm conserving pseudopotentials are used to describe the atom cores, and double-$\zeta$ polarized (DZP) atomic orbitals are taken as basis sets.\citep{Tro1991,Soler2002} GGA with the form of PBE is selected as the exchange-correlation potential. The NEGF-DFT self-consistent Hamiltonian is obtained when monitored quantities, such as every element in the Hamiltonians and the density matrices, differ by less than $1\times 10^{-4}$ eV between two iteration steps. The calculation of photocurrent involves two steps in order: (1) the Hamiltonian of the two-lead device without photons is obtained self-consistently by the NEGF-DFT formalism; (2) the electron-photon (e-p) interaction is perturbatively included as self energy during the photocurrent calculation.

Photocurrent $I_{ph}$ flowing from the right lead to the central region can be written in terms of the NEGF in the following form \citep{Chen2012,Zhang2014,Xie2015}
\begin{equation}
I_{ph}=\frac{ie}{h}\int\\Tr\Bigg\{\Gamma_{R}[G^{<}+f_{R}(G^>-G^<)]\Bigg\}dE,
\end{equation}
where $e$ is the electron charge and $h$ is the Plank constant, respectively. $\Gamma_{R}$ is the line width function of the right lead, and $f_R$ is its Fermi-Dirac distribution function. $G^{>/<}$ is the greater/lesser Green's function, which is expressed as\cite{Hen2002}
\begin{equation}
G^{>/<}=\frac{\hbar}{c\epsilon_0}\sqrt{\frac{\mu_r}{\epsilon_r}}\frac{J_n}{2\omega}
\left(\frac{e}{m_0}\right)^2~G^r_0(P^\dagger_{\theta,\rm{i}} G^{>/<}_0P_{\theta,\rm{i}})G^a_0,\label{lesser}
\end{equation}
where $c$ is the light velocity, $\hbar$ is the reduced Plank constant, and $\epsilon_0$ is the vacuum dielectric constant. $\mu_{r}$ and $\epsilon_{r}$ are the relative magnetic susceptibility and the relative dielectric constant, respectively. $J_n$ is the light intensity defined as the number of photons per unit time per unit area. $m_0$ is the static electron mass. $G_{0}^{r/a}$ and $G_{0}^{>/<}$ are the retarded/advanced and greater/lesser Green's functions without photons, respectively. $P^\dagger_{\theta,\rm{i}} G^{>/<}_0P_{\theta,\rm{i}}$ originates from the self energies of e-p interactions. The matrix $P_{\theta,{\rm i}}$ denotes the projection of the electron momentum on the polarization direction $\theta$ and atom site ${\rm i}$ which is perpendicularly irradiated by PL. For concise expression, we omit atom site ${\rm i}$ in the following content. For linearly polarized light (LPL) or elliptically polarized light (EPL), $P_{\theta} = P_{x} \cos\theta + P_{y} \sin\theta$ or $P_{x} \cos\theta \pm i P_{y} \sin\theta$, where $P_{x/y}$ is the cartesian component of the electron momentum. Substituting the expression of $P_{\theta}$, we have
\begin{equation}
\begin{split}
P^\dagger_\theta G^{>/<}_0 P_\theta=&\cos^2\theta P^\dagger_x G^{>/<}_0 P_x +\sin^{2}\theta P^\dagger_y G^{>/<}_0 P_y \\
+&\sin\theta\cos\theta (P^\dagger_x G^{>/<}_0 P_y +P^\dagger_y G^{>/<}_0 P_x )
\end{split}\label{PGP1}
\end{equation}
for LPL, and
\begin{equation}
\begin{split}
P^\dagger_\theta G^{>/<}_0 P_\theta=&\cos^2\theta P^\dagger_x G^{>/<}_0 P_x +\sin^{2}\theta P^\dagger_y G^{>/<}_0 P_y \\
\pm i&\sin\theta\cos\theta (P^\dagger_x G^{>/<}_0 P_y - P^\dagger_y G^{>/<}_0 P_x )
\end{split}\label{PGP2}
\end{equation}
for EPL, respectively. The quadratic terms ``$P_{x}P_{x}$" and ``$P_{y}P_{y}$" with the factor $\cos(2\theta)$ are identical for LPL and EPL, whereas the cross terms ``$P_{x}P_{y}$" and ``$P_{y}P_{x}$" labeled by $\sin(2\theta)$ are different for LPL and EPL. Polarization angle $\theta$ of the incident light is labeled in Fig.\ref{fig2}(a).

In order to investigate the photoelectric response properties, we define the normalized photocurrent,\citep{Hen2002,Chen2012} i.e., the photoresponse function $J_{ph} \equiv I_{ph} / eJ_n$. It is easy to show that $J_{ph}$ has the dimension of area. In our calculation, the unit of length is Bohr radius $a_0$, and hence $J_{ph}$ is measured in the unit of $a_{0}^{2}$. Suppose the photon energy $\hbar\omega\sim 1$ eV, and the optical energy flux density $J_u\sim 1W/\mu m^2$, the photoresponse function $J_{ph} \sim 1$ should be equivalent to the photocurrent $I_{ph}\sim 0.01 ~\mu A$.

In our MBP-based homojunction device, SIS is completely broken by the PN junction, which leads to the greatly enhanced photocurrent. But there is approximate mirror symmetry along the $x$-direction, since the acceptor atom Si and donor atom S are approximately located long the straight line in $x$-direction. As a result, $J_{ph}$ is nearly symmetric at $\theta = 0^{\circ}$. These results suggests that, the cross term of $J_{ph}$ labeled by $\sin(2\theta)$ is negligible, i.e., the $\sin(2\theta)$ and $\sin(-2\theta)$ terms are approximately zero; the photocurrent is mainly determined by the square terms ``$P_{x}P_{x}$" and ``$P_{y}P_{y}$". Consequently, according to Eqs.(\ref{PGP1}) and (\ref{PGP2}), photocurrents induced by LPL or EPL exhibit similar behaviors. $J_{ph}$ depends mostly on the square terms related to $\cos(2\theta)$, and hence it is nearly symmetric about $\theta=0^\circ$ and $\theta=90^\circ$. These analysis will be numerically verified in the following section.

\section{Numerical results and discussion}

\begin{figure}[tbp]
\centering
\includegraphics[width=7cm,clip=]{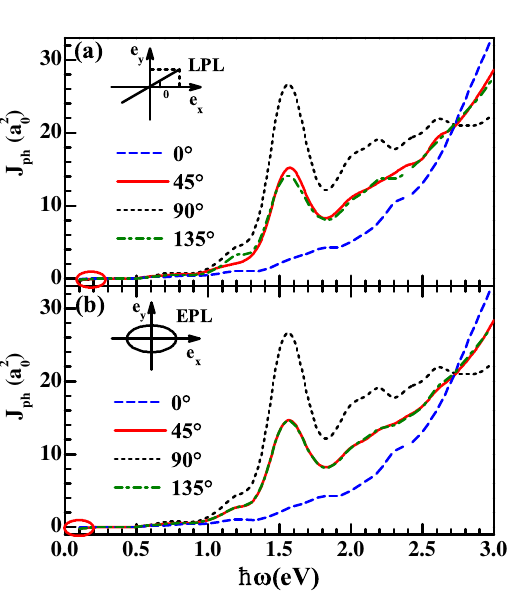}
\caption{The photocurrent versus photon energies under the irradiation of (a) linearly polarized light and (b) elliptically polarized light at different polarization angles. The polarization angle $\theta$ of the incident light is shown in the inset of panel (a).}\label{fig2}
\end{figure}

In this section, we present simulation results on the photoelectric properties of the MBP homojunction device. We first investigate the dependence of photocurrent on photon energy $\hbar\omega$ and polarization angle $\theta$. Fig.\ref{fig2} shows the photocurrent $J_{ph}$ versus $\hbar\omega$ for different $\theta$, where LPL and EPL induced photocurrents are shown in panels (a) and (b), respectively. It is obvious that the photocurrents for LPL and EPL show similar behaviors, which agrees with our analysis in Sec.{\color{blue}II}. Since $J_{ph}$ is nearly symmetric about $\theta=0^\circ$ and $\theta=90^\circ$, we see from Fig.\ref{fig2} that $J_{ph}(\theta=45^\circ) \simeq J_{ph}(\theta=135^\circ)$. Considering the similarity of LPL and EPL induced photocurrents, in the rest of the content, we focus on the photoresponse of EPL. Fig.\ref{fig2}(b) shows that, the behavior of $J_{ph}$ is distinct in different photon energy ranges. We have the following observations: (1) $J_{ph}$ is nearly zero when $\hbar\omega<0.95$ eV, which is reasonable for a semiconductor; (2) $J_{ph}$ shows a peak around $\hbar\omega\simeq 1.6$ eV for $\theta \neq 0$; (3) with the increasing of $\theta$ from $0^\circ$ to $90^\circ$, $J_{ph}$ increases when $\hbar\omega<2.7$ eV and decreases when $\hbar\omega>2.7$ eV; (4) When $\hbar\omega<0.18$ eV, photocurrent flows in the opposite direction, i.e., $J_{ph}<0$, as shown by the red ovals in Fig.\ref{fig2}.

\begin{figure}[tbp]
\centering
\includegraphics[width=\columnwidth]{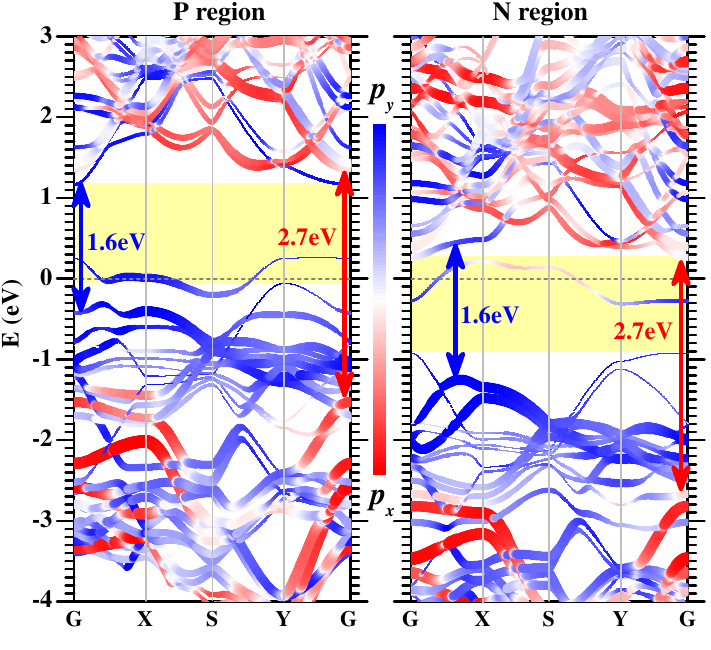}
\caption{Projected band structures for the P and N regions. The thickness of curves denotes the total contribution of $p_x$ and $p_y$ orbitals, and the color shows the relative proportion of $p_x$ and $p_y$. The blue (red) double-headed block arrows indicate possible transitions from the valence to the conduction bands, which correspond to photon energy $1.6(2.7)$ eV. The light-yellow regions represent bandgaps, and the gray dash lines correspond to the Fermi energies.}\label{fig3}
\end{figure}

Next, we analyze the above phenomena through the projected band structures for the P and N regions, which are plotted in Fig.\ref{fig3}. The thickness of curves denotes the total contribution of orbitals $p_x$ and $p_y$, and the red and blue colors denote the relative proportion of $p_x$ and $p_y$. Due to the wrinkle structure of MBP, its in-plane orbitals $p_{x}$ and $p_{y}$ widely contribute to the whole band structure, which leads to orbital-resolved photoelectric phenomenon. In the left lead of the homojunction (P region), phosphorus is partially substituted by Si, where negatively charged acceptor ions accumulate and the Fermi level is lowered. In contrast, Fermi level in the right lead (N region) is lifted. When coupled with each other, the Fermi energies of P and N regions are balanced at $E=0$ eV, at which the hybrid bands with contribution from the dopant atoms also reside. Fig.\ref{fig3} shows that the Fermi energy crossing the impurity level of the P/N region is closer to the valence/conduction band, since the band structure of the P/N region is shifted up/down. In Fig.\ref{fig3}, the energy gap is marked by the light yellow region, from which we can estimate that the relative shift is about 0.95 eV. This means the potential barrier induced by charge accumulation near the PN junction is approximately 0.95 eV. To generate a nonzero photocurrent, the charge carriers must overcome this barrier. Therefore, $J_{ph}$ significantly increases when $\hbar\omega>0.95$ eV, as shown in Fig.\ref{fig2}.

\begin{figure}
\centering
\includegraphics[width=8cm,clip=]{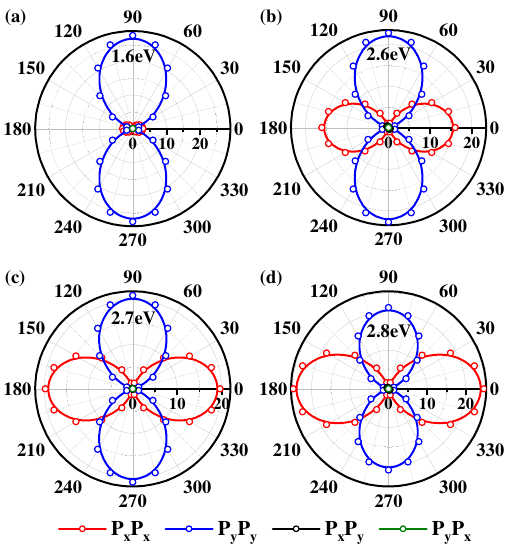}
\caption{Angle-resolved distribution of the square and cross terms of the photocurrent for different photon energies of EPL. From panel (a) to (d), the photon energy $\hbar \omega$ is 1.6 eV, 2.6 eV, 2.7 eV, and 2.8 eV, respectively. }\label{fig4}
\end{figure}

Photon-excited $e-h$ pairs contribute to the photocurrent carriers, where $e-h$ pairs are composed of holes in the valence band and electrons in the conduction band. From Fig.\ref{fig3}, we can find the curves near the top of valence band are blue ($p_y$), which means the in-plane orbital $p_y$ is dominant for the holes here. Meanwhile, the curves far below the valence band top as well as those near the conduction band bottom are red and blue ($p_x$+$p_y$), which indicates that the carriers in these bands possess nearly equal component of $p_x$ and $p_y$. According to the quantum selection rule, the electrons and holes in $e-h$ pairs must have the same orbital. In Fig.\ref{fig3}, we mark the possible direct transitions from the valence to the conduction bands, i.e., $p_y\rightarrow p_y$ (blue block arrows) and $p_x\rightarrow p_x$ (red block arrows). Photon energy $\hbar\omega=2.7$ eV is a critical value, which determines the transition is dominated by $p_y\rightarrow p_y$ (below 2.7 eV) or $p_x\rightarrow p_x$ (above 2.7 eV). On the premise of the selection rule, the more the orbital element, the larger the photocurrent. For $\hbar\omega=1.6$ eV (blue arrows), the $p_y$ element is dominant for the holes near the valence band top, and thus $J_{ph}$ reaches the local maximum for $\theta > 0$, as shown in Fig.\ref{fig2}.

Due to the $e-p$ interaction, the photocurrent is related to the electron momentum ($P_{x}$ and $P_{y}$) through the polarization angle of PL. Considering that the electron moment $P_{x/y}$ directly reflects the orbital extension of PL, the photocurrent contributed by different orbitals would have different polarizing angular dependence. Based on these facts, we plot the polarizing angular distribution for the cross and square terms of the photocurrent in Fig.\ref{fig4}. Several typical photon energies are selected: $\hbar\omega=1.6$ eV ($p_y$-based $J_{ph}$ is maximum), $\hbar\omega<2.7$ eV, $\hbar\omega=2.7$ eV (proportion boundary of $p_x$ and $p_y$), and $\hbar\omega>2.7$ eV. As expected, the cross terms ``$P_{x}P_{y}$'' and ``$P_{y}P_{x}$'' are negligible in all panels due to the mirror symmetry along $x$-direction. The square term ``$P_{x}P_{x}$'' extends along $x$-direction ($\theta=0^\circ$), while ``$P_{y}P_{y}$'' spreads along $y$-direction ($\theta=90^\circ$). For $\hbar\omega=1.6$ eV, the ``$P_{x}P_{x}$'' term is near zero, and $J_{ph}$ reaches maximum at $\theta=90^\circ$. When $\hbar\omega<2.7$ eV, the ``$P_{y}P_{y}$'' term prevails, and $J_{ph}$ is in favor of $\theta=90^\circ$. For $\hbar\omega=2.7$ eV, the magnitude of ``$P_{x}P_{x}$'' and ``$P_{y}P_{y}$'' terms are almost equal, and hence $J_{ph}$ is independent of $\theta$. Since the PN junction is infinitely extended in $x$-direction, $J_{ph}$ favors orbital $p_x$. Consequently, when $\hbar\omega>2.7$ eV, ``$P_{x}P_{x}$'' is always greater than ``$P_{y}P_{y}$'', as shown in Fig.\ref{fig4}(d). Hence $J_{ph}$ is larger at $\theta=0^\circ$ when $\hbar\omega>2.7$ eV (Fig.\ref{fig2}) due to the dominant $P_{x}P_{x}$ term extends in $x$-direction. All these facts corresponds exactly to those shown in Fig.\ref{fig2}.

\begin{figure}[tbp]
\centering
\includegraphics[width=7.5cm,clip=]{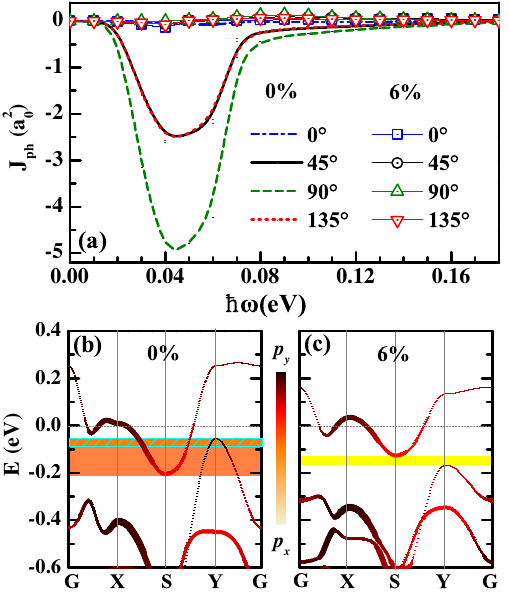}
\caption{(a): the photocurrent versus small photon energy for different tension and polarization angles. The thick lines and the symbolized lines label $J_{ph}$ of the PN junction with tension $\mathcal{T}=0\%$ and $\mathcal{T}=6\%$, respectively. (b) and (c): band structures of the left lead with tension $\mathcal{T}=0\%$ and $6\%$, respectively. }\label{fig5}
\end{figure}

We notice that in the red oval regions of Fig.~\ref{fig2}, the photocurrent flows in the opposite directions, i.e., $J_{ph}<0$ for $\hbar\omega<0.18$ eV. The photocurrent in this range is magnified in Fig.~\ref{fig5}(a). We also plot $J_{ph}$ of strained PN junction with tension $\mathcal{T}=6\%$, in which $J_{ph}$ is nearly zero at small $\hbar\omega$. Negative $J_{ph}$ is clearly observed when $\mathcal{T}=0\%$. Driven by the built-in field, $J_{ph}$ should flow from right to left, as displayed in Fig.~\ref{fig1}(c). The only possible reason for the reserved photocurrent is that charge carriers are changed from electron to hole, i.e., band inversion occurs. To interpret this abnormal phenomenon, we plot the band structures of the left lead (P region) with tension $\mathcal{T}=0\%$ and $6\%$ in Fig.~\ref{fig5}(b) and Fig.~\ref{fig5}(c), respectively.

We see that the valence band and the hybrid level of dopant atoms crossed by the Fermi energy overlap with each other (orange region in Fig.~\ref{fig5}(b)), leading to the inverse band structure. Through calculation, we find that when $\mathcal{T}$ increases, the overlap region disappears gradually, and a bandgap emerges. Fig.~\ref{fig5}(c) shows the band structure of the left lead with tension $\mathcal{T}=6\%$. This gapped band structure contrasts with the inverse band structure at $\mathcal{T}=0\%$. Comparing $J_{ph}$ in the case of $\mathcal{T}=0\%$ and $6\%$, we can deduce that the inverse band structure induces the reversed photocurrent at small $\hbar\omega$. Furthermore, in the range of small photon energy, the photocurrent is mainly contributed by orbital $p_y$. If the photon can excite carriers with maximum portion of $p_y$, PVE will be most efficient and $J_{ph}$ will show a peak. The efficient excitation occurs around  $\hbar\omega=0.045$ eV, which is marked by narrow green gridding in the orange region of Fig.~\ref{fig5}(b), where orbital $p_y$ dominates the overlapped band. As a result, $J_{ph}$ reaches the maximum at $\hbar\omega=0.045$ eV in Fig.\ref{fig5}(a).

\section{Conclusions}

In summary, we have investigated the photoelectric characteristic of a 2D lateral PN homojunction based on monolayer black phosphorus (MBP), which is driven by the photovoltaic effect under the illumination of polarized light (PL). Once the potential barrier near the PN junction is overcome, the excited photocurrent is greatly enhanced by the built-in electric field. The MBP-based PN homojunction device shows high-efficient photovoltaic response in a large photon energy range, from 1 eV to 3 eV. Due to the electron-photon ($e-p$) interaction, electron momentum of the photocurrent is closely related to the in-plane atomic orbitals of MBP. Orbital $p_{y}$ or $p_{x}$ dominates the photoelectric response in different photon energy ranges, which leads to the distinct behavior of the photocurrent for different polarization angles of the incident light. These findings indicate that the MBP-based PN homojunction has promising applications in high-efficiency photoelectric devices and orbital-resolved photovoltaic detection.

\section*{acknowledgments}
This work was supported by the National Natural Science Foundation of China (Grants No.12174023 and No.12174262).

\end{document}